\documentclass[aps,prl,twocolumn,twoside,a4paper,10pt,amsmath,amssymb,superscriptaddress]{revtex4-1}
\usepackage[switch]{lineno}
\usepackage[english]{babel}

\usepackage{mathrsfs}
\usepackage{graphicx}
\usepackage{amsmath}
\usepackage{amssymb}
\usepackage{amsfonts}
\usepackage{color}
\usepackage{leftidx}
\usepackage{verbatim}
\def\hatd#1{\hat{#1}^\dagger}
\def\bra#1{\left\langle{#1}\right|}
\def\ket#1{\left|{#1}\right\rangle}
\def\braket#1#2{\left\langle{{#1}}\mathrel{\left|{\vphantom{{#1}{#2}}}\right.\kern-\nulldelimiterspace}{{#2}}\right\rangle}

\begin{document}

\title{Quantum Double Lock-in Amplifier}

\author{Sijie Chen}
\affiliation{Institute of Quantum Precision Measurement, State Key Laboratory of Radio Frequency Heterogeneous Integration, Shenzhen University, Shenzhen 518060, China}
\affiliation{College of Physics and Optoelectronic Engineering, Shenzhen University, Shenzhen 518060, China}
\affiliation{Laboratory of Quantum Engineering and Quantum Metrology, School of Physics and Astronomy, Sun Yat-Sen University (Zhuhai Campus), Zhuhai 519082, China}

\author{Min Zhuang}
\affiliation{Institute of Quantum Precision Measurement, State Key Laboratory of Radio Frequency Heterogeneous Integration, Shenzhen University, Shenzhen 518060, China}
\affiliation{College of Physics and Optoelectronic Engineering, Shenzhen University, Shenzhen 518060, China}
\author{Ruihuang Fang}
\affiliation{Laboratory of Quantum Engineering and Quantum Metrology, School of Physics and Astronomy, Sun Yat-Sen University (Zhuhai Campus), Zhuhai 519082, China}
\author{Yun Chen}
\affiliation{Laboratory of Quantum Engineering and Quantum Metrology, School of Physics and Astronomy, Sun Yat-Sen University (Zhuhai Campus), Zhuhai 519082, China}
\author{Chengyin Han}
\affiliation{Institute of Quantum Precision Measurement, State Key Laboratory of Radio Frequency Heterogeneous Integration, Shenzhen University, Shenzhen 518060, China}
\affiliation{College of Physics and Optoelectronic Engineering, Shenzhen University, Shenzhen 518060, China}
\author{Bo Lu}
\affiliation{Institute of Quantum Precision Measurement, State Key Laboratory of Radio Frequency Heterogeneous Integration, Shenzhen University, Shenzhen 518060, China}
\affiliation{College of Physics and Optoelectronic Engineering, Shenzhen University, Shenzhen 518060, China}
\author{Jiahao Huang}
\altaffiliation{Email: hjiahao@mail2.sysu.edu.cn, eqjiahao@gmail.com}
\affiliation{Institute of Quantum Precision Measurement, State Key Laboratory of Radio Frequency Heterogeneous Integration, Shenzhen University, Shenzhen 518060, China}
\affiliation{College of Physics and Optoelectronic Engineering, Shenzhen University, Shenzhen 518060, China}
\affiliation{Laboratory of Quantum Engineering and Quantum Metrology, School of Physics and Astronomy, Sun Yat-Sen University (Zhuhai Campus), Zhuhai 519082, China}

\author{Chaohong Lee}
\altaffiliation{Email: chleecn@szu.edu.cn, chleecn@gmail.com}
\affiliation{Institute of Quantum Precision Measurement, State Key Laboratory of Radio Frequency Heterogeneous Integration, Shenzhen University, Shenzhen 518060, China}
\affiliation{College of Physics and Optoelectronic Engineering, Shenzhen University, Shenzhen 518060, China}
\affiliation{Quantum Science Center of Guangdong-Hongkong-Macao Greater Bay Area (Guangdong), Shenzhen 518045, China}

\date{\today}

\begin{abstract}
Quantum lock-in amplifier aims to extract an alternating signal within strong noise background by using quantum strategy.
However, as the target signal usually has an unknown initial phase, it is challenging to extract complete information of amplitude, frequency and phase of a signal.
Here, to overcome this challenge, we give a general protocol for achieving a quantum double lock-in amplifier and discuss its experimental feasibility.
Our protocol is accomplished via two quantum mixers under orthogonal pulse sequences.
%
%
Combining the output signals, the complete characteristics of the target signal can be obtained.
As an example, we discuss the experimental feasibility of our quantum double lock-in amplifier via a five-level double-$\Lambda$ coherent population trapping system with $^{87}$Rb atoms, in which each $\Lambda$ structure acts as a quantum mixer and the two applied dynamical decoupling sequences take the roles of two orthogonal reference signals.
Our numerical results show that the quantum double lock-in amplifier is robust against experimental imperfections, such as finite pulse length and stochastic noise.
Our study opens an avenue for extracting complete characteristics of an alternating signal within strong noise background, which is beneficial for developing practical quantum sensing technologies.
\end{abstract}
		
\maketitle

\noindent
\textbf{INTRODUCTION}

\noindent
High-precision measurement of weak alternating signals in noise background is important for both fundamental science and practical technology.
Generally, the target signal is submerged in noise background and is hard to be detected.
To obtain a high signal-to-noise ratio, one has to decrease the effect of noise and enhance the response to the target signal.
It has been demonstrated that lock-in amplifiers can extract time-dependent alternating signals from an extreme noisy background and have been applied in various fields~\cite{AT5PC1994,GBPRL2005,MLRBM2017,AICLMR2017,ZYCS2023}.
In essence, a lock-in amplifier performs a mixing process via multiplying the input signal with a reference signal, and then applies an adjustable low-pass filter for detection.
After the filter, there is almost no contribution from signals that are not at the same frequency of the reference signal, which efficiently rejects all other frequency components.
Generally, for a time-dependent alternating signals with known initial phase, one can use typical lock-in techniques with a single reference signal to extract the characteristics of the target signal.
However, if the initial phase is unknown, it is hard to use only a single reference signal to extract the complete characteristics of the target signal, such as amplitude, frequency and initial phase.
To resolve this issue, one can use double lock-in amplifier.
A double lock-in amplifier mixes the target signal with two orthogonal reference signals.
The outputs of two mixers pass through two low-pass filters and generate two signals, which can be used to extract the complete characteristics of the target signal.

By employing the rapid-developed techniques in quantum control, quantum lock-in measurements have been demonstrated and widely used for frequency measurement~\cite{SKNature2011,MZPRX2021}, magnetic field sensing~\cite{SKNature2011}, vector light shift detection~\cite{KSPRA2021} and weak-force detection~\cite{RSNC2017}.
The key for realizing a quantum lock-in amplifier is to find quantum analogs of mixing and filtering.
Using quantum probes, these two processes can be achieved by non-commutating operations and time-evolution, respectively.
{Similar to the realization of a quantum mixer~\cite{KSPRA2021}, one may use dynamical decoupling sequences as the reference signals for achieving a quantum lock-in amplifier.}
In particular, Carr-Purcell (CP) and periodic dynamical decoupling (PDD) sequences have been widely used in various quantum lock-in amplifiers, from single-particle ones~\cite{SKNature2011} to many-body ones~\cite{MZPRX2021}.
However, these schemes only consider the target signal with a known initial phase.
If the initial phase is unknown, it is difficult to extract the complete characteristics of the target signal only by means of a single PDD or CP sequence.
In analogy to a classical double lock-in amplifier, can one develop a quantum counterpart to extract complete characteristics of a target signal?

In this article, we present a general protocol for achieving a quantum double lock-in amplifier via combing double quantum interferometry with two orthogonal periodic multi-pulse sequences.
In our protocol, each quantum interferometry with a specific periodic multi-pulse sequence can be regarded as a single quantum lock-in amplifier.
The PDD and CP sequences are analogue to the two orthogonal reference signals to realize the mixing and filtering processes via time-evolution.
Thus, one can detect the target signal by combining two output signals, in which frequency, amplitude and even initial phase of the target signal can be extracted.
%
%

Our quantum double lock-in amplifier can be realized with various quantum systems.
As an example, we illustrate how to realize with a five-level double-$\Lambda$ coherent population trapping (CPT) system of $^{87}$Rb atoms.
The five-level double-$\Lambda$ system can be divided into two $\Lambda$ system by adjusting appropriate detuning, which form the two quantum mixers for a quantum double lock-in amplifier.
In such a system, one may detect the common excited-state population to obtain the combined measurement signals for extracting the target signal.
To show the experimental feasibility, we also analyze the influences of the finite pulse length and the stochastic noise.
Our numerical results show the quantum double lock-in amplifier has good robustness against these imperfections.
Our scheme opens up a feasible way for measuring the complete characteristics of an alternating signal within strong noise background.

\begin{figure*}[!htp]
\includegraphics[width=2\columnwidth]{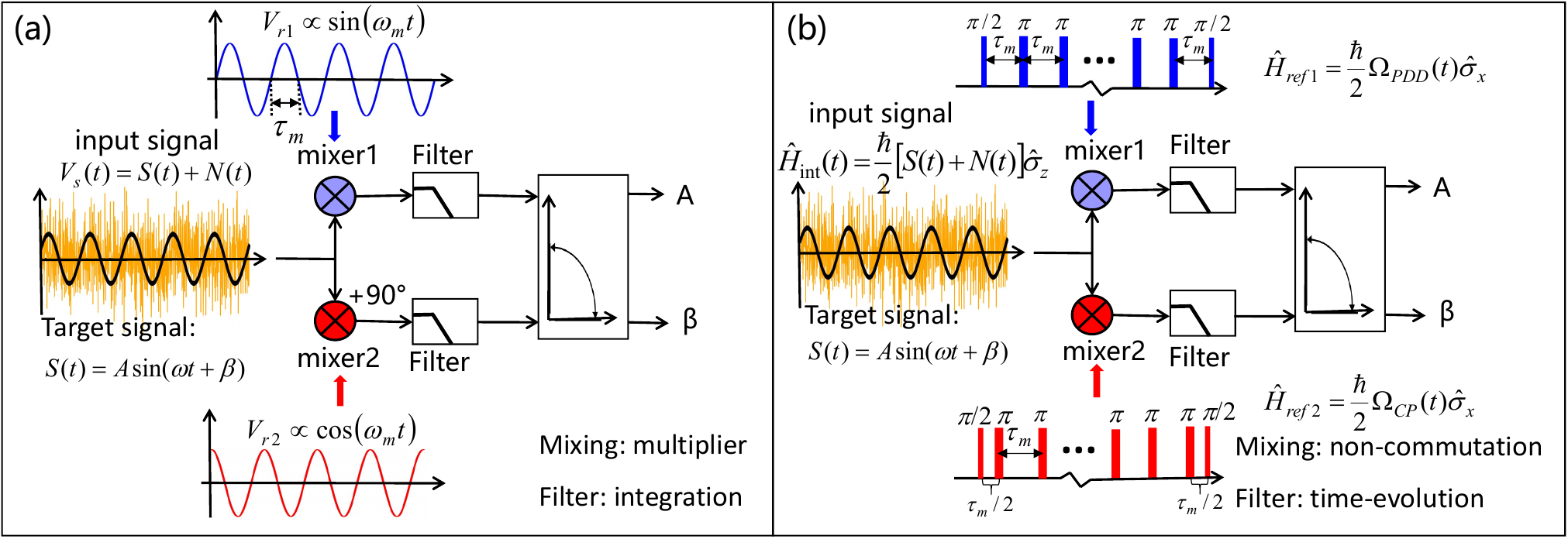}
\caption{\label{Fig1}(color online).
The schematic of classical and quantum double lock-in amplifiers.
(a) The classical double lock-in amplifier.
$V_{s}(t)=\emph{S}(t)+\emph{N}(t)$ is the input signal, where $\emph{S}(t)=A\sin(\omega t+\beta)$ is the target signal submerged within the noise $\emph{N}(t)$.
$V_{r1,2}(t)$ are the two orthogonal reference signals.
The amplitude $A$, frequency $\omega$, and phase $\beta$ can be extracted after mixing with a multiplier and filtering by integration.
(b) The quantum double lock-in amplifier.
There are two identical quantum mixers.
For each quantum mixer, the coupling between the probe and the signal is described by ${\hat{H}}_\textrm{int}=\frac{\hbar}{2}\emph{M}(t) \hat{\sigma}_{z}$ where $\emph{M}(t)=\emph{S}(t)+\emph{N}(t)$ includes the target signal $\emph{S}(t)$ and the noise $\emph{N}(t)$.
The mixing modulations ${\hat{H}}_\textrm{ref1}=\frac{\hbar}{2}\Omega_\textrm{PDD}(t) \hat{\sigma}_{x}$ and ${\hat{H}}_\textrm{ref2}=\frac{\hbar}{2}\Omega_\textrm{CP}(t) \hat{\sigma}_{x}$ (implemented by the PDD and CP sequences respectively), which do not commute with ${\hat{H}}_\textrm{int}$, are analogue to the two reference signals $V_{r1}(t)$ and $V_{r2}(t)$.
Each mixer obeys the Hamiltonian ${\hat{H}}={\hat{H}}_\textrm{int}+{\hat{H}}_\textrm{ref1,2}$, which can be regarded as a single quantum lock-in amplifier.
The mixing process is achieved by non-commutating operations, and the filtering process is realized by time-evolution.
The combination of the two quantum lock-in amplifiers forms a quantum double lock-in amplifier, which can extract the complete characteristics of the target signal $\emph{S}(t)=A \sin(\omega t+\beta)$.
}
\end{figure*}

\noindent
\textbf{\\RESULTS\label{Sec2}}

\noindent
\textbf{\\General protocol}.

\noindent
In this section, we introduce the general protocol of a quantum double lock-in amplifier, which aims to extract the complete characteristics of a target signal within strong noise background.
In general, a conventional classical lock-in amplifier cannot effectively extract the phase information of the target signal.
However, a classical double lock-in amplifier can solve this problem.
By mixing the input signal $\emph{V}_s(t)=\emph{S}(t)+\emph{N}(t)$ with two orthogonal reference signals $\emph{V}_{r1}(t)=\sin(\omega_m t)$ and $\emph{V}_{r2}(t)=\cos(\omega_m t)$ respectively and integrating the two mixed signals over a certain time, the target signal can be extracted, see Fig.~\ref{Fig1}~(a).
Here, $\emph{S}(t)= A\sin(\omega t+\beta)$ is the target signal submerged in the noise $\emph{N}(t)$ and all three parameters $(A,\omega,\beta)$ are unknown to be measured and unchanged in measurements.
%
%
%
The two multipliers, which are described by $\textrm{V}_{r1}^\textrm{mix}(t)=\emph{V}_s(t)\times\emph{V}_{r1}(t)$ and $\textrm{V}_{r2}^\textrm{mix}(t)=\emph{V}_s(t)\times\emph{V}_{r2}(t)$, are used for mixing input and reference signals.
The integrator is used to filter out the components whose frequencies are different from the reference frequency $\omega_m$.
One can find that, at the lock-in point $\omega_m=\omega$, the two output signals are given as $I=\frac{AT}{2}\cos(\beta)$ and $Q=\frac{AT}{2}\sin(\beta)$.
%
Therefore, at the lock-in point, one can obtain $A=2\sqrt{I^2+Q^2}/{T}$ and $\beta=\arctan(Q/I)$ for the target signal~\cite{SGIAM2015,GBAPS2008}.
%
%
In particular, if the noise spectral components are far from the reference frequency $\omega_m$, the noise effects will be averaged out through the integration (see Supplementary Note A for more details).
%
%
%

In analogy to a classical double lock-in amplifier, a quantum double lock-in amplifier can be realized by using two orthogonal multi-pulse sequences, which act the role of two orthogonal reference signals.
As the applied multi-pulse sequences (acting as the reference signal) are non-commutating with the target signal~\cite{SKNature2011,MZPRX2021,WGPRX2022}, one can mix the input signal and the reference signal.
And then the following time-evolution filters out {the noise spectral components} different from the reference frequency $\omega_m$. 

\begin{figure}[!htp]
\includegraphics[width=\columnwidth]{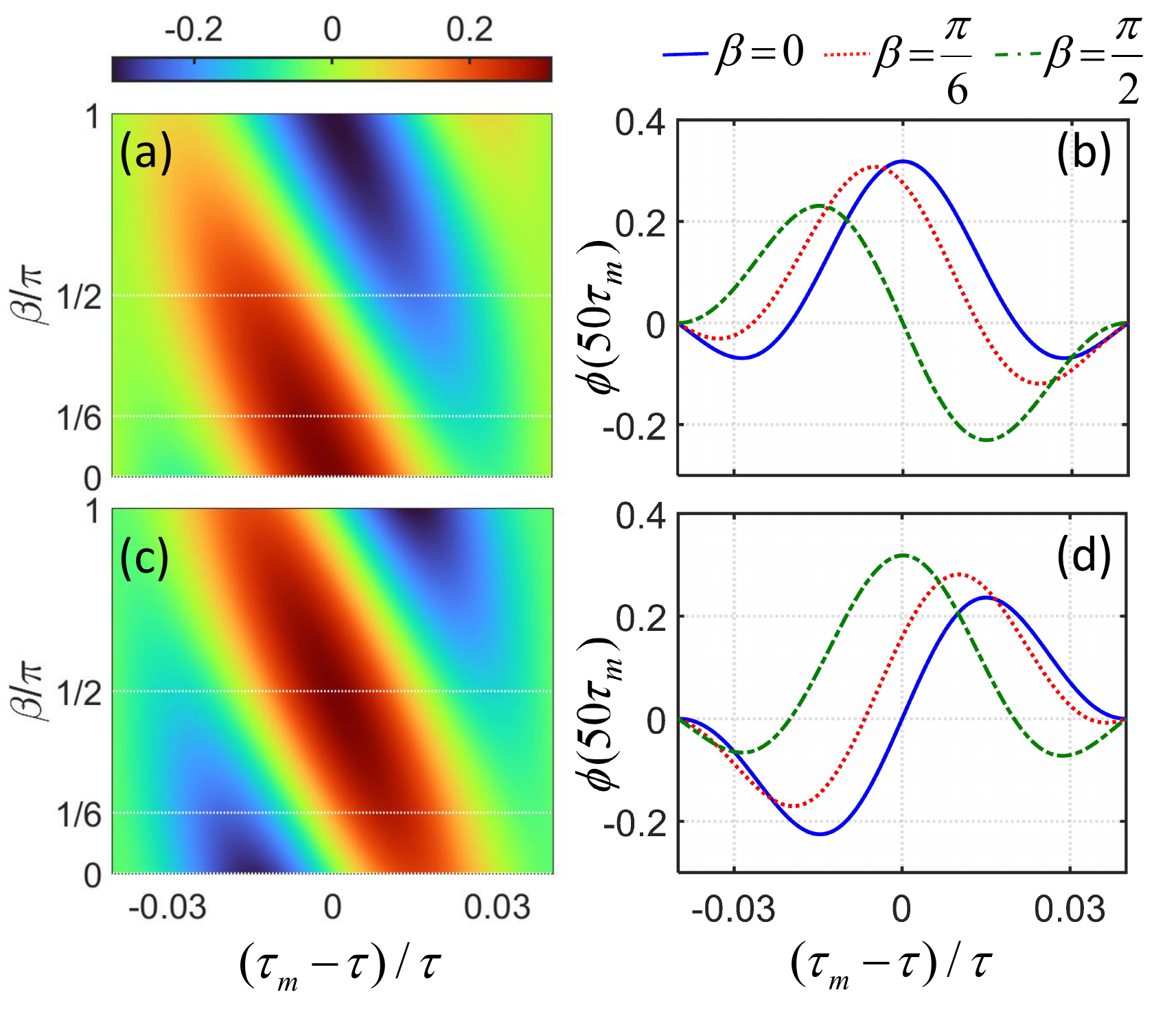}
	\caption{\label{Fig2}(color online).
		The parameter dependence of the two total accumulation phases $\phi^\textrm{PDD}_n$ and $\phi^\textrm{CP}_n$.
        (a) The variations of $\phi^\textrm{PDD}_n$ versus $(\tau-\tau_{m})$ and $\beta$.
        (b) The variations of the phase $\phi^\textrm{PDD}_n$ versus $(\tau-\tau_{m})$ with $\beta=0,\pi/6,\pi/2$.
        The phase $\phi^\textrm{PDD}_n$ is symmetric (or antisymmetric) with respect to the lock-in point $\tau=\tau_{m}$ when $\beta=0$ (or $\beta=\pi/2$),
        and is not symmetric with respect to the lock-in point $\tau=\tau_{m}$ when $\beta=\pi/6$.
		(c) The variations of $\phi^\textrm{CP}_n$ versus $(\tau-\tau_{m})$ and $\beta$.
        (d) The variations of $\phi^\textrm{CP}_n$ versus $(\tau-\tau_{m})$ with $\beta=0,\pi/6,\pi/2$.
        The phase $\phi^\textrm{CP}_n$ is antisymmetric (or symmetric) with respect to the lock-in point $\tau=\tau_{m}$ when $\beta=0$(or $\beta=\pi/2$),
        and is not symmetric (or antisymmetric) with respect to the lock-in point $\tau=\tau_{m}$ when $\beta=\pi/6$.
		Here, $A=1/100$, $\omega=\pi$ and $T=50\tau_m$.
	}
\end{figure}

To illustrate our protocol, we consider two individual two-level systems whose energy levels are labeled by $\ket{\uparrow}$ and $\ket{\downarrow}$.
For each two-level system, the coupling between the probe and the external signal is described by the Hamiltonian $\hat{H}_\textrm{int}=\frac{\hbar}{2}\emph{M}(t)\hat{\sigma}_z$ with the Pauli operators $\hat{\sigma}_{x,y,z}$.
The external signal $\emph{M}(t)=S(t)+ N(t)$ consists of the target signal $S(t)=A \sin(\omega t+\beta)$ and the stochastic noise $N(t)$.
%
%
There are different protocols for achieving mixing.
Here we consider the mixing term $\hat{H}_\textrm{ref}=\frac{\hbar}{2}\Omega (t)\hat{\sigma}_x$, which does not commute with $\hat{H}_\textrm{int}$.
Thus, the whole Hamiltonian reads
\begin{equation}\label{HS}
\hat{H}=\hat{H}_\textrm{int}+\hat{H}_\textrm{ref}=\frac{\hbar}{2}\left[M(t)\hat{\sigma}_{z}+\Omega({t})\hat{\sigma}_{x}\right].
\end{equation}
The time-evolution obeys the Schr\"{o}dinger equation,
\begin{eqnarray}\label{SE}
i\frac{\partial{\ket{\Psi(t)}_S}}{\partial t}
=\frac{1}{2} \left[M(t)\hat{\sigma}_{z}+\Omega({t})\hat{\sigma}_{x}\right] \ket{\Psi(t)}_S,
\end{eqnarray}
where $\ket{\Psi(t)}_S$ denotes the system state.
In the interaction picture with respecting to $\hat{H}_\textrm{ref}$ (See {Supplementary Note B} for more details), the time-evolution obeys
{
\begin{eqnarray}\label{IE}
i\frac{\partial{\ket{\Psi(t)}_I}}{\partial{t}}
=\frac{1}{2}M(t)\left[\cos(\alpha (t))\hat{\sigma}_{z}+\sin(\alpha(t))\hat{\sigma}_{y}\right]{\ket{\Psi(t)}_I}\nonumber\\
\end{eqnarray}}
with $\ket{\Psi(t)}_I=e^{i\int_{0}^{t}\hat{H}_\textrm{ref}(t')dt'/\hbar}\ket{\Psi (t)}_S$ and $\alpha(t)=\int^{t}_{0} \Omega(t')dt'$.
The instantaneous state at time $t$ reads
\begin{equation}\label{EITOA}
\ket{\Psi(t)}_I=\hat{\mathcal{T}}e^{-i\frac{1}{2}\left[\phi_z(t)\hat{\sigma}_z+\phi_y(t)\hat{\sigma}_y\right]}\ket{\Psi(0)}_I,
\end{equation}
where $\hat{\mathcal{T}}$ denotes the time-ordering operator, and $\phi_z(t)=\int_0^t\omega_z(t')dt'$ and $\phi_y(t)=\int_0^t\omega_y(t')dt'$ with angular frequencies $\omega_z(t)=\emph{M}(t)\cos(\alpha(t))$ and $\omega_y(t)=\emph{M}(t)\sin(\alpha(t))$.
In our scheme, the time-dependent modulation $\Omega(t)$ is designed as a sequence of $\pi$ pulses with equidistant spacings.
This technique is mature and has been widely used in quantum sensing for measuring an oscillating signal in the presence of noises~\cite{GDLPRL2011,SKNature2011,JMBPRL2016,JMBScience2017,RSNC2017,SSScience2017,MZPRX2021}.
In experiments, the $\pi$ pulse sequences can usually be approximated as square waves,
\begin{equation}\label{Os}
\Omega_\textrm{S}(t,T_{\Omega})=\left\{
\begin{array}{rl}
\pi/T_{\Omega},&|t-t_j|<T_{\Omega}/2,\\
0,&\textrm{others},
\end{array}
\right.
\end{equation}
where $T_{\Omega}$ denotes the pulse length.
In the limit of $T_{\Omega}~\to~0$, the {$\pi$-pulse} sequences can be described by hard pulses
\begin{equation}\label{PDD,CP}
\Omega(t)=\pi\sum_{j=1}^{N_p}\delta[t-(j-\lambda)\tau_m],
\end{equation}
with $\delta(t)$ being the Dirac $\delta$ function, $N_p$ denoting the pulse number, and $\tau_m$ describing the spacing of the adjacent $\pi$ pulses.
Here, the parameter $\lambda$ determines the relative phase with respect to the target signal and $\lambda=(0, 1/2)$ respectively correspond to (PDD, CP) sequences~\cite{JBNP2011,CLDRMP2017}.
{The PDD and CP sequences are two orthogonal pulse sequences, due to their initial relative phase is $\frac{\pi}{2}$}.
{For hard $\pi$ pulses, one can easily find $\alpha(t)=\int^{t}_{0} \Omega(t')dt'=N_p\pi$, therefore we have $\sin [\alpha(t)]=0$}.
Initializing each quantum system into the state $\ket{\Psi}_\textrm{in}=({\ket{\uparrow}+\ket{\downarrow}})/{\sqrt{2}}$, in the interaction picture, the instantaneous state at time $t_n=n\tau_m$ becomes
\begin{eqnarray}\label{FI}
\ket{\Psi(t_n)}_I&=&e^{-i\int_0^{t_n}\frac{1}{2}M(t')\cos[\alpha(t')]dt'\hat{\sigma}_z}\ket{\Psi}_\textrm{in},\nonumber\\
&=&(\ket{\uparrow}+e^{i\phi_n}\ket{\downarrow})/{\sqrt{2}},
\end{eqnarray}
with $n$ chosen as a even number to suppress DC noise~\cite{CLDRMP2017}.
{For the results determined by the density matrix
$\rho = \ket{\Psi}\bra{\Psi}$, the global phase $e^{-i\phi_n/2}$ can be ignored.}
Back to the Schr\"{o}dinger picture~\cite{SKNature2011,MZPRX2021}, the output state is
\begin{eqnarray}\label{FS}
\ket{\Psi(t_n)}_S
&=&({\ket{\uparrow}+e^{i(-1)^{N_p}\phi_n}\ket{\downarrow}})/{\sqrt{2}},
\end{eqnarray}
with the pulse number $N_p$. 
For PDD and CP sequences, the total accumulation phases are   %
\begin{eqnarray}\label{phi_PDD}
\phi^\textrm{PDD}_n&=&\frac{2A}{\omega}\cos\left[\frac{n\omega\cdot(\tau_m-\tau)}{2}+\beta\right]\\\nonumber
&\times& \cos\left[\frac{\omega\cdot(\tau_m-\tau)}{2}\right]\frac{\sin\left[n\omega\cdot(\tau_m-\tau)/2\right]}{\sin\left[\omega\cdot(\tau_m-\tau)/2\right]},
\end{eqnarray}
and
\begin{eqnarray}\label{phi_CP}
\phi^\textrm{CP}_n&=&\frac{2A}{\omega}\sin\left[\frac{n\omega\cdot(\tau_m-\tau)}{2}+\beta\right]\\\nonumber
&\times&\left[1+\sin\left(\frac{\omega\cdot(\tau_m-\tau)}{2}\right)\right]\frac{\sin[n\omega\cdot(\tau_m-\tau)/2]}{\sin[\omega\cdot(\tau_m-\tau)/2]},
\end{eqnarray}
respectively (see Supplementary Note B for more details).
Here $\tau={\pi}/{\omega}$ is the half period of the target signal.
Obviously, given $\beta=0$ {($\beta=\pi/2$)}, $\phi^\textrm{PDD}_n$ is symmetric (anti-symmetric) and $\phi^\textrm{CP}_n$ is anti-symmetric (symmetric) with respect to the lock-in point $\tau_m=\tau$, see Fig.~\ref{Fig2}.

In the stage of signal extraction, an unitary operation $U=e^{-i\frac{\pi}{4}{\hat{\sigma}_y}}$ is applied for recombination and the readout state becomes
$\ket{\Psi}_\textrm{re}=\left[i\sin(\frac{\phi_n'}{2})\ket{\uparrow}+\cos(\frac{\phi_n'}{2})\ket{\downarrow}\right]$ with $\phi_n'=(-1)^{N_p}\phi_n$.
Hence the final probability of the probe in the state $\ket{\uparrow}$ are
$P_{\uparrow,n}^\textrm{PDD}={[1-\cos(\phi^\textrm{PDD}_n)]}/2$
and
$P_{\uparrow,n}^\textrm{CP}={[1-\cos(\phi^\textrm{CP}_n)]}/2$
respectively.
The corresponding expectations of $z$-component Pauli operator are
$\langle\hat{\sigma}_z\rangle_{n}^\textrm{PDD}=-{\cos(\phi^\textrm{PDD}_n)}$
and
$\langle\hat{\sigma}_z\rangle_{n}^\textrm{CP}=-{\cos(\phi^\textrm{CP}_n)}$
respectively.
Thus through modulating the pulse repetition period $\tau_m$, one can determine the lock-in point from the pattern symmetry, and the amplitude can be extracted from Eq.~\eqref{phi_PDD} or Eq.~\eqref{phi_CP} via a fitting procedure~\cite{SKNature2011,MZPRX2021,CLDRMP2017,WGPRX2022,JMNC2021}.
However, when $\beta \neq 0$ or $\beta \neq \pi/2$ , the symmetry of $\phi^\textrm{PDD(CP)}_n$, $P^\textrm{PDD(CP)}_{\uparrow,n}$ and $\langle\hat{\sigma}_z\rangle_{n}^\textrm{PDD(CP)}$ are destroyed and one cannot determine the lock-in point from the spectra, see Fig.~\ref{Fig2}.
This means that one cannot extract the complete characteristics of the target signal $\emph{S}(t)$ only by means of a single PDD or CP sequence.
Below, we introduce how to solve this issue via the quantum double lock-in amplifier.
For convenience, we divide the target signals into two types: (i) the weak signals of $\frac{A}{\omega}\leq\frac{1}{2n}$ and (ii) the strong signals of $\frac{A}{\omega}>\frac{1}{2n}$.

\begin{figure}[!htp]
\includegraphics[width=\columnwidth]{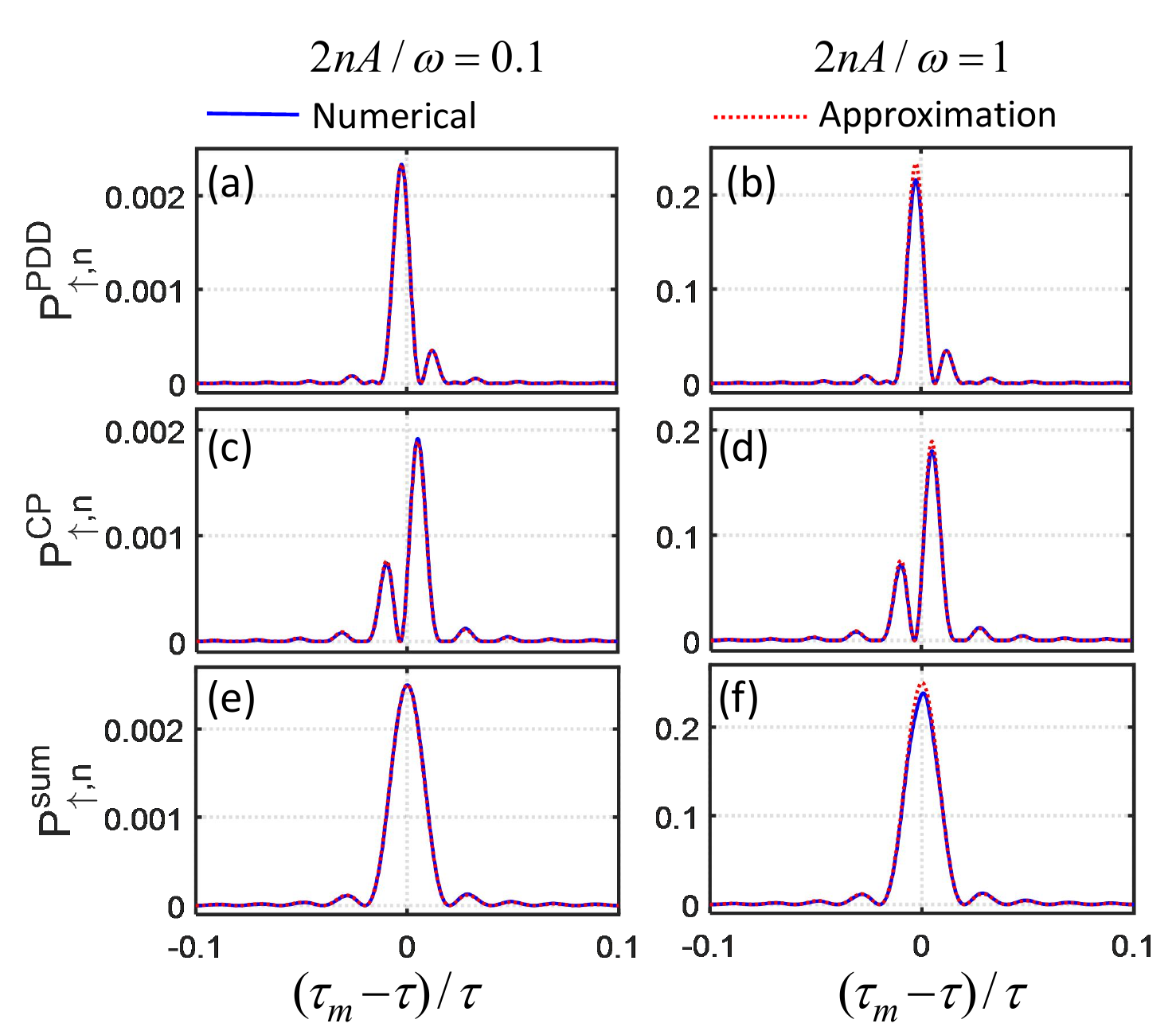}
	\caption{\label{Fig3}(color online).
		Extraction of a weak target signal.
        (a)(b)The variations of the measurement signals $P_{\uparrow,n}^\textrm{sum}$ versus $(\tau_{m}-\tau)$.
        The measurement signals $P_{\uparrow,n}^\textrm{sum}$  is symmetric with respect to the lock-in point $\tau_m=\tau$ and it is well consistent with the analytically approximate result Eq.~\eqref{Psum}.
        (c)(d)The variations of the measurement signals $P_{\uparrow,n}^\textrm{PDD}$ versus $(\tau_{m}-\tau)$.
        The measurement signals $P_{\uparrow,n}^\textrm{PDD}$ is not symmetric with respect to the lock-in point $\tau_m=\tau$ and it is well consistent with the analytically approximate result.
        (e)(f)The variations of the measurement signals $P_{\uparrow,n}^\textrm{CP}$ versus $(\tau_m-\tau)$.
        The measurement signals $P_{\uparrow,n}^\textrm{CP}$ is not symmetric with respect to the lock-in point $\tau_m=\tau$ and it is well consistent with the analytically approximate result.
        %
        Here, we choose $2nA/\omega=0.1$ [left: (a), (c) and (e)] and the critical case $2nA/\omega=1$ [right: (b), (d)and (f)] with $n=100$, $\beta=-\pi/6$ and $\omega = \pi$.
	}
\end{figure}
%

For weak signals, i.e. $\frac{A}{\omega}\leq\frac{1}{2n}$, we choose the sum of $P_{\uparrow,n}^{PDD}$ and $P_{\uparrow,n}^{CP}$ as a measurement signal to recover the spectrum symmetry, that is,
%
\begin{eqnarray}\label{Psum}
P_{\uparrow,n}^\textrm{sum}&=&P_{\uparrow,n}^\textrm{PDD}+P_{\uparrow,n}^\textrm{CP}\\\nonumber
&\approx&\left(\frac{A}{\omega}\right)^2\left[\frac{\sin[n\omega\cdot(\tau_m-\tau)/2]}{\sin[\omega\cdot(\tau_m-\tau)/2]}\right]^2.
\end{eqnarray}
Obviously, $P_{\uparrow,n}^\textrm{sum}$ is symmetric with respect to the lock-in point $\tau_m=\tau$ again, see Fig.~\ref{Fig3}~(e, f).
Thus, one can determine the lock-in point from the symmetric pattern of $P_{\uparrow,n}^\textrm{sum}$ versus $(\tau_m-\tau)$, which can be obtained by adjusting the spacing $\tau_m$ of adjacent $\pi$ pulses.
Our analytical results are well consistent with the numerical ones, even for the case of $\frac{A}{\omega}=\frac{1}{2n}$.
Once the value of $\omega$ is determined via the the lock-in point $\tau_m=\tau$, one can extract $A$ from Eq.\eqref{Psum} via a fitting procedure.
Moreover, due to the value of $\omega$ and $A$ are both extracted, one can determine $\beta$ from $P_{\uparrow,n}^\textrm{PDD}$ or $P_{\uparrow,n}^\textrm{CP}$.
Thus the complete characteristics of the target signal can be obtained within our scheme.

For strong signals, i.e. $\frac{A}{\omega}>\frac{1}{2n}$, we choose the sume of $\langle\hat{\sigma}_z\rangle_{n}^\textrm{PDD}$ and $\langle\hat{\sigma}_z\rangle_{n}^\textrm{CP}$ as a measurement signal, that is,
\begin{eqnarray}\label{Szsum}
\langle\hat{\sigma}_z\rangle_{n}^\textrm{sum}&=&\langle\hat{\sigma}_z\rangle_{n}^\textrm{PDD} + \langle\hat{\sigma}_z\rangle_{n}^\textrm{CP}\nonumber \\
&=&-\left[{\cos(\phi^\textrm{PDD}_n)+\cos(\phi^\textrm{CP}_n)}\right].
\end{eqnarray}
At the lock-in point $\tau_m=\tau$, it reads
\begin{small}
\begin{eqnarray}\label{Szsum1}
\langle\hat{\sigma}_z\rangle_{n,\tau_{m}=\tau}^\textrm{sum}&=&-\cos\left[\frac{2 A}{\omega}\cos(\beta)\cdot n\right]-\cos\left[\frac{2 A}{\omega}\sin(\beta)\cdot n\right], \nonumber \\
\end{eqnarray}
\end{small}
which is an exactly bisinusoidal oscillation.
By using a fast Fourier transform (FFT), one can determine the lock-in point by judging whether $\langle\hat{\sigma}_z\rangle_{n}^\textrm{sum}$ has bisinusoidal oscillatory pattern.
%
%
%
In comparison to $P_{\uparrow,n}^\textrm{sum}$, the FFT of $\langle\hat{\sigma}_z\rangle_{n}^\textrm{sum}$ does not contain the zero-frequency component and it is easy to determine the lock-in point from the FFT.
\begin{figure}[!htp]
\includegraphics[width=\columnwidth]{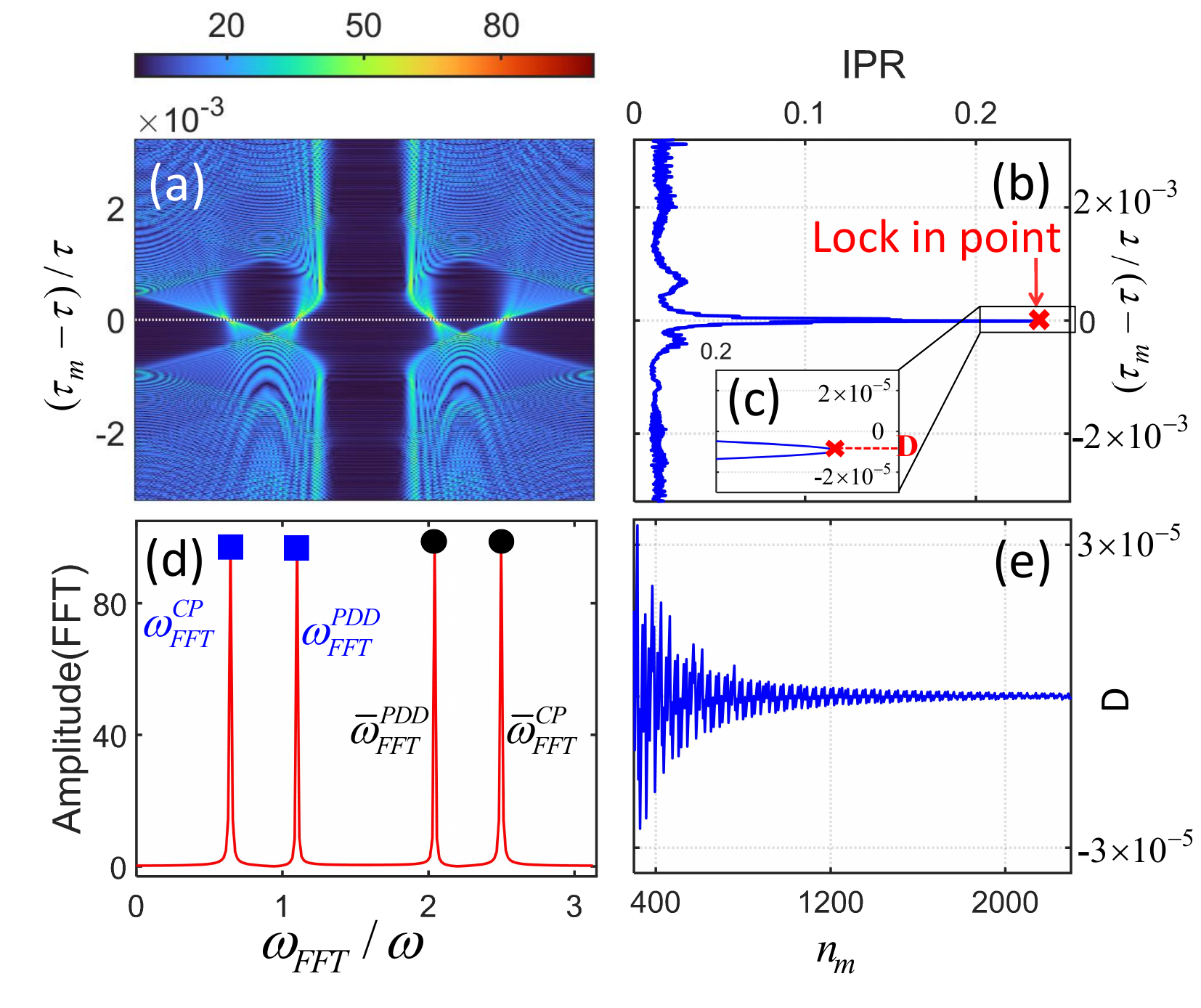}
	\caption{\label{Fig4}(color online).
		Extraction of a strong target signal.
		(a) The FFT spectrum of $\langle\hat{\sigma}_z\rangle_{n}^\textrm{sum}$ versus $(\tau_m-\tau)$ with even positive integers $n$ up to $n_m=400$.
        Given $\tau=\tau_m$, the FFT spectrum just has four peaks and one can use it to determine the lock-in point.
        %
		(b) The IPR versus $(\tau_m-\tau)$.
         The maximum of $\textrm{IPR}$ can be used to determine the lock-in point.
        (c) The inset for the local amplification region in (b) and denotes the shift of lock-in point $D$.
		(d) The FFT of $\langle\hat{\sigma}_z\rangle_{n}^\textrm{sum}$ at $\tau_m=\tau$.
         The four peaks locate at $\omega_\textrm{FFT}^\textrm{CP}/\omega=\frac{2}{\omega}A|\sin(\beta)|=0.637$, $\omega_\textrm{FFT}^\textrm{PDD}/\omega=\frac{2}{\omega}A|\cos(\beta)|=1.103$, $\overline\omega_\textrm{FFT}^\textrm{PDD}/\omega=(\pi-\omega_\textrm{FFT}^\textrm{PDD}/\omega)$ and $\overline\omega_\textrm{FFT}^\textrm{CP}/\omega=(\pi-\omega_\textrm{FFT}^\textrm{CP}/\omega)$ respectively.
		(e) The variation of the shift $D$ versus the maximum sensing scanning time $n_m$.
		Here, we choose $A=2$, $\beta=-\pi/6$ and $\omega=\pi$.
	}
\end{figure}
In our analysis, we consider a series of measurement results $\langle\hat{\sigma}_z\rangle_{n}^\textrm{sum}$ for different evolution times $t_n=n\tau_m$.
As shown in Fig.~\ref{Fig4}~(a), we give the FFT spectrum of $\langle\hat{\sigma}_z\rangle_{n}^\textrm{sum}$ versus $\tau_m$.
{When $\tau_m=\tau$, the FFT spectrum has only four peaks and therefore we can determine the lock-in point via its inverse participation ratio (IPR)~\cite{MCJSM2015,NCMPRB2011,FEPRL2000,TBPCIOP2018,MYCBIJ2018}, which is defined as
\begin{equation}\label{IPR}
\textrm{IPR}=\frac{\sum_{k=2,even}^{n_m}|F_k|^4}{\left|\sum_{k=2,even}^{n_m}|F_k|^2\right|^2},
\end{equation}
with
%
$F_k=\sum_{n=2,\textrm{even}}^{n_m}\langle\hat{\sigma}_z\rangle_{n}^\textrm{sum}e^{-i\left(\frac{2\pi}{n_m}nk\right)}$
%
and $t_{n_m}=n_m\tau_m$ is the maximum sensing time.
}
After some algebra, when $\omega(\tau_m-\tau) \ll 1$ and $n_{m} \to \infty$, {we have ${\textrm{IPR}={1}/{4}}$ for $\tau_m=\tau$ and ${\textrm{IPR}=0}$ for $\tau_m\neq\tau$} (see Supplementary Note C for more details).
%
%
Given $\tau_m=\tau$, the four peaks appear at $\omega_\textrm{FFT}^\textrm{CP}=2A|\sin(\beta)|$, $\omega_\textrm{FFT}^\textrm{PDD}=2A|\cos(\beta)|$, $\overline\omega_\textrm{FFT}^\textrm{PDD}=(\pi\omega-\omega_\textrm{FFT}^\textrm{PDD})$ and $\overline\omega_\textrm{FFT}^\textrm{CP}=(\pi\omega-\omega_\textrm{FFT}^\textrm{CP})$ respectively, see Fig.~\ref{Fig4}(c).
In which, the two peaks at $\omega_\textrm{FFT}^\textrm{CP}$ and $\omega_\textrm{FFT}^\textrm{PDD}$ correspond to the two oscillation frequencies of the measurement signal $\langle\hat{\sigma}_z\rangle_{n}^\textrm{sum}|_{\tau_{m}=\tau}$.
Thus, the values of $A$ and $\beta$ are given as
\begin{equation}\label{S0}
A=\frac{1}{2}\sqrt{(\omega_\textrm{FFT}^\textrm{PDD})^2+(\omega_\textrm{FFT}^\textrm{CP})^2}
\end{equation}
and
\begin{equation}\label{beta}
|\beta|=\arctan\left({\omega_\textrm{FFT}^\textrm{CP}}/{\omega_\textrm{FFT}^\textrm{PDD}}\right),
\end{equation}
which is similar to a classical double lock-in amplifier.
Moreover, one can determine the sign of parameters $\beta$ from $\langle\hat{\sigma}_z\rangle_{n}^\textrm{PDD}$  or $\langle\hat{\sigma}_z\rangle_{n}^\textrm{CP}$ via a fitting procedure.
Considering the coherence time of a quantum system is limited in experiments, we numerically calculate the systems with moderate $n_{m}$.
{It is shown that $\textrm{IPR}\approx1/4$ for $\tau_m=\tau$ and $\textrm{IPR}\approx0$ for $\tau_m\neq\tau$}, thus we can still extract the frequency $\omega$ by finding the maximum of $\textrm{IPR}$, see Fig.~\ref{Fig4}~(b).
Further, given the measurement signal $\langle\hat{\sigma}_z\rangle_{n}^\textrm{sum}|_{\tau_{m}=\tau}$, we can extract $A$ and $\beta$ via Eq.~\eqref{S0} and Eq.~\eqref{beta}, see Fig.~\ref{Fig4}~(d).

In experiments, the coherence time $T_2$ is finite and thus the value of $n_m$ should satisfy $T_2>n_m\tau$.
Given the value of $n_m$, the frequency $\omega_\textrm{FFT}$ can only take discrete values $\omega_k=k\frac{2\pi\omega}{n_m}$ ($k=0,1,2,\cdots,n_m-1)$ and thus the lock-in point may be shifted.
In Fig.~\ref{Fig4}~(c), we show the frequency shift $\textrm{D}$ of the lock-in point.
We numerically calculate how the frequency shift $\textrm{D}$ varies with $n_{m}$ and find that the envelop of $\textrm{D}$ decreases with $n_m$ due to the resolution ratio of FFT frequency $\omega_\textrm{FFT}$ is $(\omega_{k+1}-\omega_k)=\frac{2\pi\omega}{n_m}\propto1/n_m$, see Fig.~\ref{Fig4}~(e).
%

\noindent
\textbf{\\Experimental feasibility.}

\begin{figure}[!htp]
\includegraphics[width=\columnwidth]{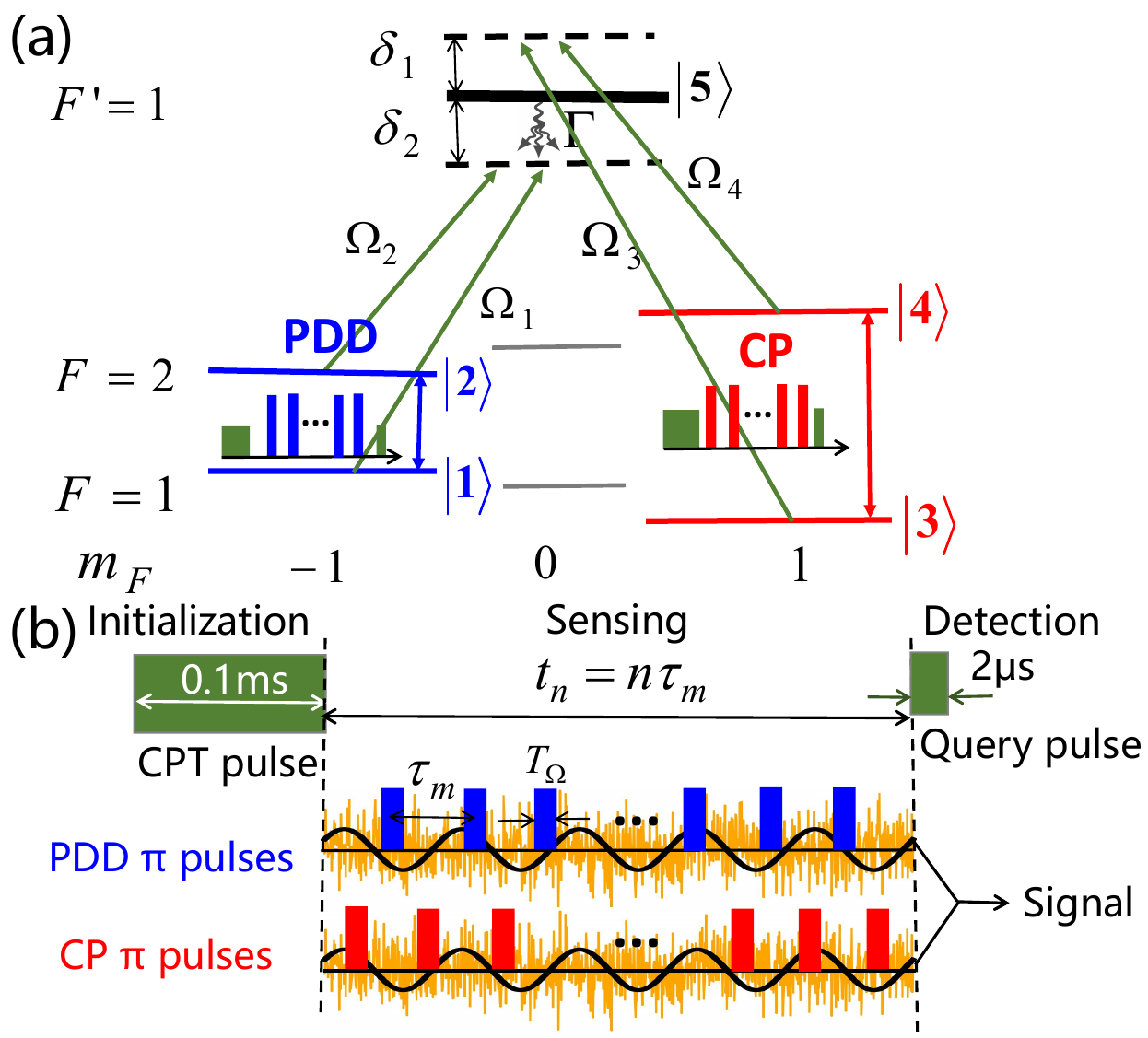}
	\caption{\label{Fig5}(color online).
		Schematic of the quantum double lock-in amplifier via five-level double-$\Lambda$ CPT in $^{87}$Rb.
		(a) Five-level double-$\Lambda$ configuration of $^{87}$Rb atom which includes $\ket{1}=\ket{F=1,m_F=-1}$, $\ket{2}=\ket{F=2,m_F=-1}$, $\ket{3}=\ket{F=1,m_F=1}$, $\ket{4}=\ket{F=2,m_F=1}$ and $\ket{5}=\ket{F'=1,m_F=0}$.
		(b) Realization of the quantum double lock-in amplifier.
		Initialization: preparing the two dark states as the initial states.
		Sensing: coupling the two ground states $\{\ket{1},\ket{2}\}$ and $\{\ket{3},\ket{4}\}$ through PDD and CP sequences respectively.
         Here, $T_{\Omega}$ denotes the ${\pi}$ pulse length, $\tau_m$ is the pulse repetition periods and the total sensing time is $t_n=n\tau_m$.
		 Detection: a CPT pulse with $2$~$\mu$s is imposed to detect the common excited state population via fluorescence or transmission spectrum.
	}
\end{figure}

\noindent
Our quantum double lock-in amplifier can be realized via a double-$\Lambda$ coherent population trapping~(CPT) system, in which each $\Lambda$ system is employed as a quantum mixter.
As shown in Fig.~\ref{Fig5}~(a), the double-$\Lambda$ CPT system can be realized by simultaneously coupling two independent two-level systems ($\{\ket{1},\ket{2}\}$ and $\{\ket{3},\ket{4}\}$) with a common excited state ($\ket{5}$), which has been extensively observed in alkali atoms such as Cs~\cite{GBPRL2005,PYJP2016,MAHJAP2017} and Rb~\cite{JVAPB2005,RFNPJ2021,LMPRA2012}.
Moreover, such a system can be divided into two single-$\Lambda$ systems by splitting the Zeeman sublevels of their ground states~\cite{RFNPJ2021}, which provide the two independent physical channels.
The simultaneous coupling of these two physical channels can be achieved by $lin||lin$ CPT scheme, in which two CPT light fields are linearly polarized to the same direction and orthogonal to the applied magnetic field~\cite{RFNPJ2021,ZAWD2017,EEMOS2010}.

In general, for a single-$\Lambda$ system, the CPT lasers will pump the atoms into a dark state which is a coherent superposition of two ground states~\cite{YFAPB2015,ZAWD2017,MAHJAP2017}.
To achieve our quantum double lock-in amplifier, we can simultaneously prepare two dark states as the initial states of two physical channels by a pulse which contains four CPT frequency components.
After the initialization via CPT pulse, the prepared two dark states can independently accumulate phases under the control of PDD and CP sequences, respectively.
During the signal interrogation process, the applied PDD and CP sequences only couple the ground states in the same quantum mixer, see Fig.~\ref{Fig5}~(a).
Then through applying the CPT procedure, the coherent property of ground states becomes proportional to the population of the common excited state, which can be experimentally detected by fluorescence or transmission spectrum.
Therefore, the accumulated phases in our quantum double lock-in amplifier can be obtained via a CPT query pulse that simultaneously couples the two physical channels.

In the following, we show an example of quantum double lock-in amplifier via a five-level double-$\Lambda$ system constructed by the $D1$ line of $^{87}$Rb atoms.
The involved energy levels include two groups of magneto-sensitive states $\{\ket{1}=\ket{F=1,m_F=-1},\ket{2}=\ket{F=2,m_F=-1}\}$ and $\{\ket{3}=\ket{F=1,m_F=1},\ket{4}=\ket{F=2,m_F=1}\}$ coupling with a common excited state $\ket{5}=\ket{F'=1,m_F=0}$, as shown in Fig.\ref{Fig5}~(a).
Their eigenfrequencies are respectively labeled as $\omega_{j}$ $(j=1,2,3,4,5)$.
The total decay rate from the excited state ($\ket{5}$) to the four ground states is $\Gamma = 2\pi \times 5.746$ MHz.
To perform the simultaneous coupling, the CPT pulse should contain four frequency components $(\omega_{L1},\omega_{L2},\omega_{L3},\omega_{L4})$ with Rabi frequencies $(\Omega_1,\Omega_2,\Omega_3,\Omega_4)$, which can be generated by modulating a single laser with a fiber-coupled electro-optic phase modulator (EOPM).
For convenience, one can set the four Rabi frequencies $\Omega_j=\Omega$ are real and the four decay rates $\gamma_j=\Gamma/4$~$(j=1,2,3,4)$~\cite{ZAWD2017,RFNPJ2021}.
The time-evolution obeys the Lindblad master equation~\cite{ZAWD2017},
\begin{equation}\label{Lindblad}
\frac{\partial\hat{\rho}}{\partial t}=-\frac{i}{\hbar}\left[\hat{H},\hat{\rho}\right]
+\sum_{j=1}^{4} \frac{\Gamma}{4}\left(\hat{L}_j\hat{\rho}\hat{L}_j^{\dagger}-\frac{1}{2}\hatd{L}_j\hat{L}_j\hat{\rho}-\frac{1}{2}\hat{\rho}\hatd{L}_j\hat{L}_j\right),
\end{equation}
where $\hat{L}_j=\ket{j}\bra{5}$ is the Lindblad operator, and $\hat{\rho}$ is the density matrix.
In the rotating frame, the Hamiltonian matrix is given as
{
\begin{equation}\label{Heff}
	\hat{H}\!=\!\hbar\!\left(
	\begin{array}{ccccc}
	-\delta_1\!-\!\frac{\Delta_1}{2} &0 \!&0 \!&0 \!&\Omega^*\\
	0&-\delta_1\!+\!\frac{\Delta_1}{2}\!&0 \!&0 \!&\Omega^*\\
	0 \!&0 \!&\delta_2\!-\!\frac{\Delta_2}{2} \!&0 \!&\Omega^*\\
	0 \!&0 \!&0\!&\delta_2\!+\!\frac{\Delta_2}{2} \!&\Omega^*\\
	\Omega \!&\Omega \!&\Omega \!&\Omega \!&0
	\end{array}\right),
\end{equation}
}
where $\Delta_1=(\omega_2+\omega_{L2})-(\omega_1+\omega_{L1})$ and $\Delta_2=(\omega_4+\omega_{L4})-(\omega_3+\omega_{L3})$ are the two-photon detunings,  and $\delta_1=[\omega_5-(\omega_2+\omega_{L2}+\omega_1+\omega_{L1})/2]$ and $\delta_2=[(\omega_4+\omega_{L4}+\omega_3+\omega_{L3})/2-\omega_5]$ are the average detunings.

In order to realize the quantum double lock-in amplifier, we input two dark states $\ket{D_{12}}={(\ket{1}-\ket{2})}/{\sqrt{2}}$ and
$\ket{D_{34}}={(\ket{3}-\ket{4})}/{\sqrt{2}}$ as two probe states which can be prepared via CPT procedure.
In experiments, one can prepare the two dark states via setting a big gap between the average detunings $\delta_1$ and $\delta_2$ which satisfy the far detuning condition: $\delta_2=\delta_1\gg\{\Omega,\Delta_{1,2}\}$, see Fig.~\ref{Fig5}~(b).
Thus the corresponding density matrix reads $\hat{\rho}_D=\frac{1}{2}\left(\ket{D_{12}}\bra{D_{12}}+\ket{D_{34}}\bra{D_{34}}\right)$.
During the signal interrogation process, due to the two physical channels have different resonance frequencies and there is a strong bias field magnetic field $B_{bias}$, the two dark states $\ket{D_{12}}$ and $\ket{D_{34}}$ will independently accumulate phases $\phi_n^\textrm{PDD}$ and $\phi_n^\textrm{CP}$ under the corresponding PDD and CP sequences (see Supplementary Note D for more details).

In our calculations, the two physical channels $\{\ket{1},\ket{2}\}$ and $\{\ket{3},\ket{4}\}$ are decoupled in the whole process due to the far detuning condition, and the detunings are set as $\delta_1=\delta_2=2\pi\times1$ MHz.
%
%
At time $t_n=n\tau_m$ before detection, the density matrix is $\hat{\rho}_f=\left(\ket{\Psi_{12}(t_n)}\bra{\Psi_{12}(t_n)}+\ket{\Psi_{34}(t_n)}\bra{\Psi_{34}(t_n)}\right)/2$ with
$\ket{\Psi_{12}(t_n)}=({\ket{1}-e^{i\phi^\textrm{PDD}_n}\ket{2}})/{\sqrt{2}}$ and  $\ket{\Psi_{34}(t_n)}=({\ket{3}-e^{i\phi^\textrm{CP}_n}\ket{4}})/{\sqrt{2}}$.
%
At last, a CPT query pulse of $2$ $\mu$s is imposed to obtain the population of the common excited state.
Here, the transmission of CPT light can be converted as the transmission signal (TS) via a photodetector~\cite{RFNPJ2021,ZAWD2017}, which is proportional to $(1- \rho_{55})$, i.e. the absorption is proportional to the excited-state population $\rho_{55}$.
That is, the readout is not directly measuring the population $\rho_{55,n}$ from the coherence $\rho_{12,n}$ and $\rho_{34,n}$ (see Supplementary Note D for more details).
For a weak signal, we have the common excited-state population
\begin{eqnarray}\label{rho55(0)}
\rho_{55,n}&\approx&\frac{\Omega^2}{\Gamma^2}\left[1+2\Re(\rho_{12,n})+2\Re(\rho_{34,n})\right]\\\nonumber
&\approx&\frac{\Omega^2}{\Gamma^2}\left(\frac{A}{\omega}\right)^2\left[\frac{\sin[n\omega\cdot(\tau_m-\tau)/2]}{\sin[\omega\cdot(\tau_m-\tau)/2]}\right]^2,
\end{eqnarray}
with the density matrix element $\rho_{ij,n}$ at time $t_n=n\tau_m$ being proportional to the measurement signal $P_{\uparrow,n}^\textrm{sum}$ given in Eq.~\eqref{Psum}.
According to Eq.\eqref{rho55(0)}, one can extract the weak signal from the common exicted-state population $\rho_{55,n}$, which is proportional to the CPT light absorption and can be detected via a photodetector~\cite{RFNPJ2021,GSPOSA2011}.
For a strong signal, we define
\begin{eqnarray}\label{Drho55}
\tilde\rho_{55,n}&=&\rho_{55,n}-\frac{1}{(n_m/2)}\sum_{n'=2,\textrm{even}}^{n_m}\rho_{55,n'}\\\nonumber
&\approx&-\frac{\Omega^2}{2\Gamma^2}\left[\cos(\phi_n^\textrm{PDD})+\cos(\phi_n^\textrm{CP})\right]
\end{eqnarray}
which is proportional to $\langle\hat{\sigma}_z\rangle_n^\textrm{sum}$ given in Eq.~\eqref{Szsum}.
{At the point of $\tau_m=\tau$, we have $\tilde\rho_{55,n}\approx -\frac{\Omega^2}{2\Gamma^2}\left[\cos\left(\frac{2A}{\omega}\cos(\beta)\cdot n\right)+\cos\left(\frac{2A}{\omega}\sin(\beta)\cdot n\right)\right]$,  which can be used to extract the strong signal (see Supplementary Note D for more details)}.

Therefore, according to Eqs.~(19) and (20), one can successfully extract the target signal from the population of the common excited-state in a five-level double-$\Lambda$ CPT system.
{In experiment, one can obtain the common excited-state population $\rho_{55,n}$ and further obtain $\tilde\rho_{55,n}$ by adjusting the spacing $\tau_m$ of adjacent $\pi$ pulses.}
In our consideration, we chose $\{\ket{1},\ket{2}\}$ and $\{\ket{3},\ket{4}\}$ to form two groups of magneto-sensitive transitions with gyromagnetic ratios $\gamma_{12}=-\gamma_{34}=\gamma_{g}=-(g_2-g_1)\mu_B/\hbar=-1.0014\times2\pi\times1.4$ $\textrm{MHz/G}$, where $\mu_B$ is the Bohr magneton, $\{g_1,g_2\}$ are the Land$\acute{e}$ $g$ factors for ground states $F=\{1,2\}$, and $\mu_B/\hbar=2\pi\times1.4$ $\textrm{MHz/G}$ and $(g_2-g_1)=1.0014$.
Below, for a target AC magnetic field in form of $B_0\sin(\omega t+\beta)$, we use $A=\gamma_g B_0$ to denote its amplitude.
Based upon currently available experimental techniques, we set the frequency $\omega=2\pi\times50$ kHz, the initial phase $\beta=-\pi/6$, the $\pi$ pulse length $T_{\Omega}=2$ $\mu$s and the Rabi frequencies $\Omega=0.035\times\frac{\Gamma}{4}$ for simulation.
To split the Zeeman sublevels, a bias magnetic field $B_{bias}=0.143$ mT is applied while the two-photon detunings are set as $\Delta_{1}=\Delta_2\approx0$, leading to the average detunings $\delta_{1}=\delta_2\approx2\pi\times1$ MHz.

\begin{figure}[!htp]
\includegraphics[width=\columnwidth]{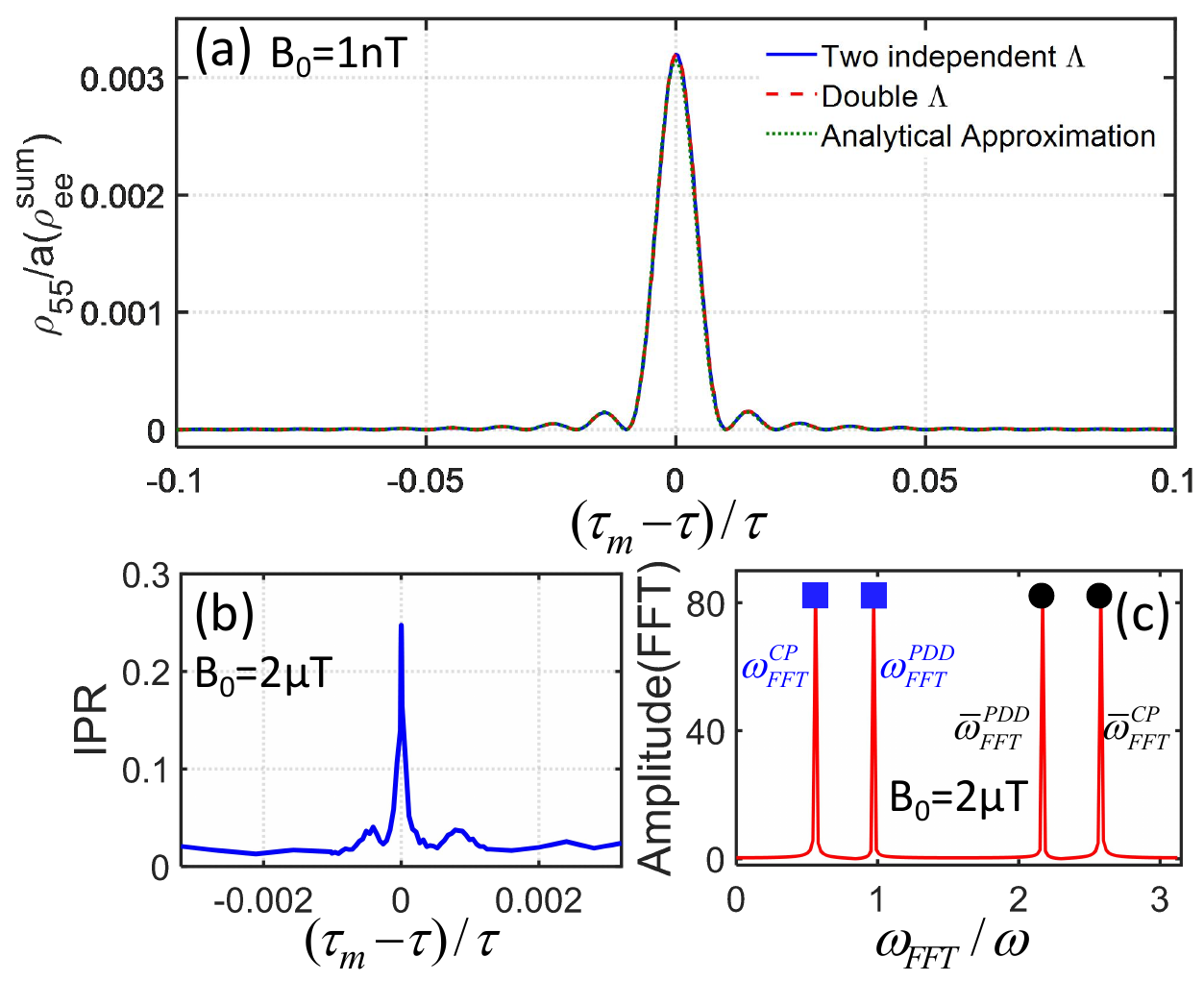}
	\caption{\label{Fig6}(color online).
		Numerical results of the quantum double lock-in amplifier via five-level double-$\Lambda$ CPT in $^{87}$Rb.
        (a) Locking of a weak target signal via the normalized common excited-state population $\rho_{55}$ versus $(\tau_m-\tau)$.
        The numerical results of $\rho_{55}/a$ (dashed red line) fit well with the sum of excited-state populations given by two independent $\Lambda$ configurations (solid blue line), and they are both consistent with the analytical approximation (dotted green line).
         Here, $B_0=1$ nT, $n=200$, $\omega=2\pi \times 50$ kHz ($\tau=10$ $\mu$s), $\beta=-\pi/6$, and $T_{\Omega}=2$ $\mu$s.
		(b) Locking of a strong target signal via the \textrm{IPR} versus $(\tau_m-\tau)$.
		The \textrm{IPR} approach its maximum at the lock-in point $\tau_m=\tau$.
		(c) The FFT spectrum of $\tilde\rho_{55,n}$ for $\tau_m=\tau$.
        The first two peaks are $0.587$ and $0.968$ which are very close to the theoretical ones $\omega_\textrm{FFT}^\textrm{CP}/\omega=2A|\sin(\beta)|/\omega=0.561$
        and $\omega_\textrm{FFT}^\textrm{PDD}/\omega=2A|\cos(\beta)|/\omega=0.971$.
       Here, $B_0=2$ $\mu$T, $n=2,4,\cdots,400$, $\omega=2\pi \times 50$ kHz($\tau=10$ $\mu$s), $\beta=-\pi/6$, and $T_{\Omega}=2$ $\mu$s.
	}
\end{figure}

Our numerical results from the Lindblad master equation~\eqref{Lindblad} are well consistent with the analytical ones given by Eqs.~\eqref{rho55(0)} and \eqref{Drho55}, see Fig.~\ref{Fig6}.
For a weak target signal, such as a magnetic field of $B_0=2\times 10^{-9}$ T satisfying $\frac{A}{\omega}\leq\frac{1}{2n}$,
one can obtain the information of the target signal by measuring $\rho_{55,n}$, see Fig.~\ref{Fig6}~(a).
Based upon the numerical simulation of Lindblad master equations, the normalized common excited-state population ($\rho_{55}/a$ with the normalization coefficient $a\approx\frac{\Omega^2}{\Gamma^2}$) given by a five-level double-$\Lambda$ system (red dashed line) fit well with the sum of excited-state populations given by two independent $\Lambda$ systems (blue line).
Moreover, these numerical results are both well consistent with the analytical ones of Eq.(19) (green dotted line).
{Thus one can determine the lock-in point to extract the frequency $\omega$ from the common excited-state population.}
In addition, one may also determine $\beta$ from $P_{\uparrow,n}^\textrm{PDD}$ or $P_{\uparrow,n}^\textrm{CP}$ via additional experiments, which are just applied PDD or CP sequences on a single-$\Lambda$ system.
%
For a strong target signal, such as a magnetic field $B_0=2\times 10^{-6}$ T satisfying $\frac{A}{\omega}>\frac{1}{2n}$, one can determine the value of $\omega$ via the $\textrm{IPR}$ given by the FFT spectra of the measurement signal $\tilde\rho_{55,n}$, see in Fig.~\ref{Fig6}~(b).
Therefore one can extract $A$ and $\beta$ from the FFT spectra of $\tilde\rho_{55,n}$ at the lock-in point $\tau_m=\tau$, see Fig.~\ref{Fig6}~(c).
Similarly, we find the numerical results are well consistent with the analytical ones, see Fig.~\ref{Fig6}~(b) and (c).
The little deviation of numerical between analytical results is because that the pulse is not a ideal Dirac $\delta$ function but a square pulse of a finite length $T_{\Omega}=2$ $\mu $s.

\noindent
\textbf{\\Robustness.}
%
	
\begin{figure}[!htp]
	\includegraphics[width=\columnwidth]{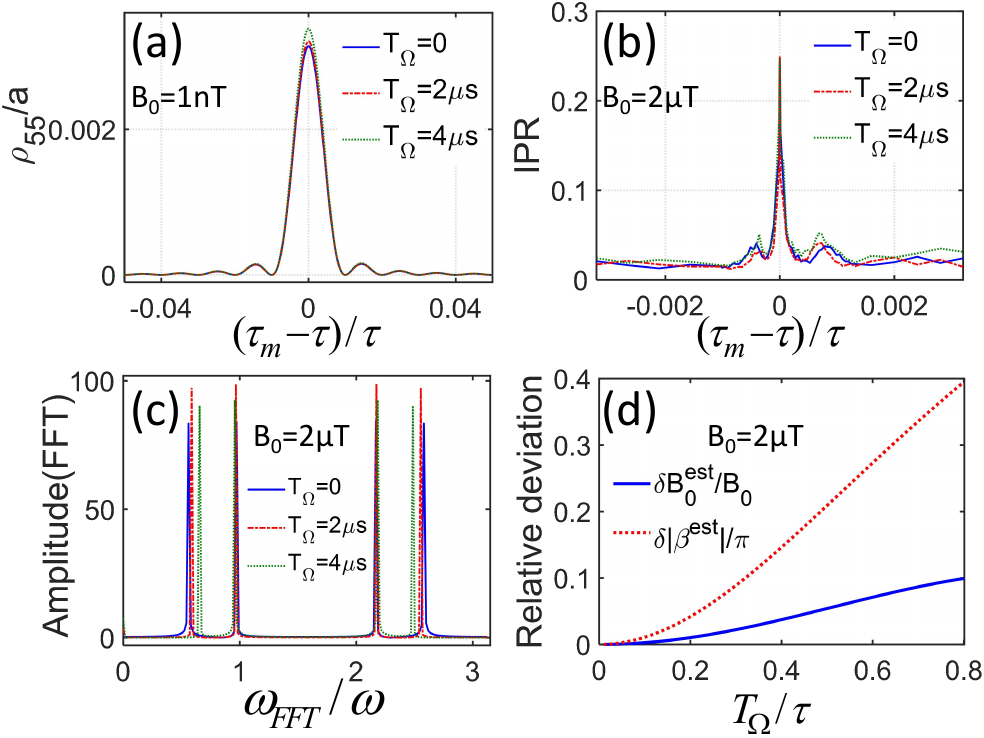}
	\caption{\label{Fig7}(color online).
		The influence of finite pulse length on the quantum double lock-in amplifier.
         (a) For the weak target signal measurement, the normalized excited state population $\rho_{55}/a$ versus $(\tau_m-\tau)$ with different pulse length $T_{\Omega}=0$ (solid blue line), $T_{\Omega}=2$ $\mu$s (dashed red line) and $T_{\Omega}=4$ $\mu$s (dotted green line).
        It indicates that the pulse length does not affect the lock-in point when $T_{\Omega}\leq0.4\tau$.
         Here,$B_0=1$ nT, $\omega=2\pi\times50$ kHz ($\tau=10$ $\mu$s), $\beta=-\pi/6$, $n=200 (n\tau=2$ ms).
         (b) For the strong target signal measurement, the \textrm{IPR} versus $(\tau_m-\tau)$ with different pulse length $T_{\Omega}=0$ (solid blue line), $T_{\Omega}=2$ $\mu$s (dashed red line) and $T_{\Omega}=4$ $\mu$s (dotted green line).
        It also indicates that the pulse length does not affect the lock-in point.
		(c) The FFT results of $\tilde\rho_{55,n}$ in the case of $\tau_m=\tau$ with different pulse length $T_{\Omega}=0$ (solid blue line), $T_{\Omega}=2$ $\mu$s (dashed red line) and $T_{\Omega}=4$ $\mu$s (dotted green line).
        (d) The relative error relative deviation of the amplitude $B_0$ and the phase $\beta$ versus the pulse length $T_{\Omega}$. The effect of pulse length $T_{\Omega}$ can be ignored if $T_{\Omega}\leq0.2\tau$.
		Here, $B_0=2$ $\mu$T, $\omega=2\pi\times50$ kHz ($\tau=10$ $\mu$s), $\beta=-\pi/6$ and $n=0,2,4,\cdots,n_m(n_m=400,n_m\tau=4$ ms).}
\end{figure}

\noindent
In experiments, there are many imperfections that may influence the lock-in signal.
Below we discuss two key imperfections: the finite pulse length $T_{\Omega}$ and the stochastic noises.
{Firstly}, we consider square pules and analyze the influence of their pulse length $T_{\Omega}$ on our scheme.
{According to Eq.~\eqref{EITOA}, for a weak target signal, we can ignore the time-ordering operator $\hat{\mathcal{T}}$, that is, $\ket{\Psi(t)}_I=e^{-i\frac{1}{2}\left[\phi_z(t)\hat{\sigma}_z+\phi_y(t)\hat{\sigma}_y\right]}\ket{\Psi(0)}_I$ with $\ket{\Psi(0)}_I=\frac{\ket{\uparrow}+\ket{\downarrow}}{\sqrt{2}}$.}
Hence, after an unitary operation $U=e^{-i\frac{\pi}{4}{\hat{\sigma}_y}}$ for readout, the population in the state $\ket{\uparrow}$ reads
\begin{equation}\label{PupOs}
P_{\uparrow,n}\approx\frac{1-\cos\left(\sqrt{\phi_y(t_n)^2+\phi_z(t_n)^2}\right)}{2}.
\end{equation}
{The analytical results of Eq.~\eqref{PupOs} are well consistent with the corresponding numerical ones (see Supplementary Note E for more details).}
To show the influences of the pulse length $T_{\Omega}$, we numerically calculate the common exited-state population $\rho_{55,n}$ in double-$\Lambda$ system with $T_{\Omega}=\{0,0.2\tau,0.4\tau\}$.
%
It indicates that the finite pulse length $T_{\Omega}$ almost has no effects on the lock-in point, see Fig.~\ref{Fig7}~(a).
For a strong target signal, {one can still} extract the frequency via the periodicity of the measurement signal $\langle\hat{\sigma}_z\rangle_{n}^\textrm{sum}$.
When $\tau_m=\tau$, we have $\hat{H}(t+2\tau_m)=\hat{H}(t)$ and the IPR reaches its maximum.
We also numerically verify that the pulse length $T_{\Omega}$ indeed does not affect the lock-in point, as shown in Fig.~\ref{Fig7}~(b).
Moreover, we numerically calculate the FFT spectra for different $T_{\Omega}$ at the lock-in point $\tau_m=\tau$, see Fig.~\ref{Fig7}~(c).
%
%
In addition, in order to estimate the influence of the pulse length $T_{\Omega}$, we numerically calculate the relative deviation of the amplitude $B_0$ and the phase $\beta$ for different $T_{\Omega}$.
They are $\delta B_0^{\textrm{est}}=\left(B_0^{\textrm{est}}-B_0\right)$ and $\delta|\beta|^\textrm{est}=\left(|\beta|^\textrm{est}-|\beta|\right)$.
In general, the pulse length need satisfy $0<T_{\Omega}<\tau$ to avoid the mixing term become a continuous driving.
Here, $B_0$ and $|\beta|$ denote the exact values, and $B_0^{\textrm{est}}$ and $|\beta|^\textrm{est}$ correspond to the estimated values according to Eq.~\eqref{S0} and Eq.~\eqref{beta}, respectively.
Our numerical results show that the effect of pulse length $T_{\Omega}$ is small enough to be ignored when $T_{\Omega}\leq0.4\tau$, as shown in Fig.~\ref{Fig7}~(d).

%

\begin{figure}[!htp]
\includegraphics[width=\columnwidth]{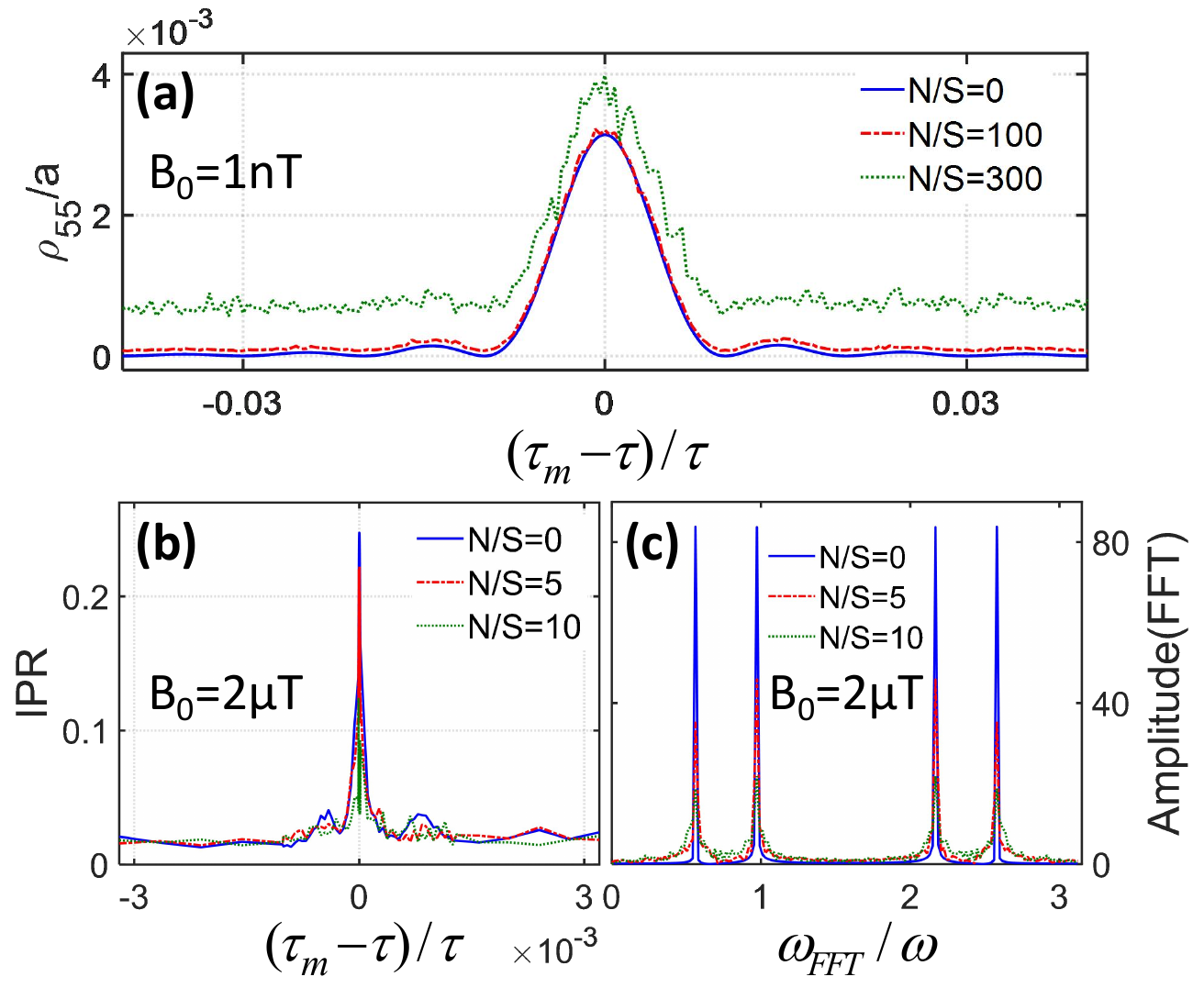}
	\caption{\label{Fig8}(color online).
	Robustness of the quantum double lock-in amplifier against white noises. The parameters for the target signal are chosen as $\omega_0=2\pi\times50$ kHz ($\tau=10$ $\mu$s) and $\beta=-\pi/6$.
     (a) For a weak target signal with $B_0=1$ nT, the normalized common excited-state population $\rho_{55}/a$ after 200 $\pi$-pulses versus $(\tau_m-\tau)$ under different SNR $N/S=0$ (solid blue line, without noise), $N/S=100$ (dashed red line) and $N/S=300$ (dotted green line).
     (b) For a strong target signal with $B_0=2$ $\mu$T, the \textrm{IPR} versus $(\tau_m-\tau)$ under different SNR $N/S=0$ (solid blue line, without noise), $N/S=5$ (dashed red line) and $N/S=10$ (dotted green line).
	(c) The FFT spectra of $\tilde\rho_{55,n}$ in the case of $\tau_m=\tau$ under different SNR $N/S=0$ (solid blue line, without noise), $N/S=5$ (dashed red line) and $N/S=10$ (dotted green line).
     In (b) and (c), the pulse number is chosen as $n=0,2,4,\cdots,400$.}
\end{figure}

Generally, stochastic noise is always exists and of course will affect the measurement precision.
Below we will illustrate the robustness of our quantum double lock-in amplifier against stochastic noise.
In experiments, one of the most common and most dominant types of noise is white noise. 
%
The white noise is a random signal equably distributed in the whole frequency domain and it is also called $f_0$ noise for its constant power spectral density. 
{
Due to the PDD and CP sequences can enhance the part with frequency $f_k=\frac{(2k+1)}{2\tau_m}(k=0,1,2,\cdots)$~\cite{CLDRMP2017}, the signal-to-noise ratio~(SNR) can be effectively improved by these two sequences.
Here, we consider the signal and the noise couple to the probe through the same channel, that is, the system obeys the Hamiltonian
$\hat{H}(t,T_{\Omega},\sigma)=\frac{\hbar}{2}\left[\emph{S}(t)+\emph{N}_{(0,\sigma)}(t)\right]\hat{\sigma}_z+\frac{\hbar}{2}\Omega(t)\hat{\sigma}_x$ with $\emph{N}_{(0,\sigma)}(t)=\sigma\cdot\emph{N}_{(0,1)}(t)$ denoting the Gaussian random noise of the time-averaged value $\overline{\emph{N}_{(0,\sigma)}(t)}=0$ and the standard deviation $\sigma=N/S$.
}
To illustrate the robustness of our scheme, we numerically calculate the two measurement signals versus the modulation period $\tau_m$ under different noise strengths.
For a weak target signal, we calculate the population $\rho_{55,n}$ versus the difference $(\tau_m-\tau)$ for different noise strengths $\sigma=0$, $\sigma=100$, and $\sigma=300$, as shown in Fig.~\ref{Fig8}~(a).
%
The stochastic noise almost do not affect the lock-in point even under a large noise strength $\sigma=100$.
Our protocol is still hold for $\textrm{SNR}\geq-20$ dB after averaging over $20$ times.
For a strong target signal, we calculate the \textrm{IPR} versus $(\tau_m-\tau)$ and the FFT spectra of $\tilde\rho_{55,n}$ for different noise strengths $\sigma=0$, $\sigma=5$, and $\sigma=10$, see Fig.~\ref{Fig8}~(b).
Our results show that the maximum IPR and the FFT amplitude decrease with the noise strength.
When $\textrm{SNR}\geq-10$ dB, one can still determine the lock-in point via the \textrm{IPR} and extract the target signal via performing FFT on $\tilde\rho_{55,n}$ (averaging over $20$ times), see Fig.~\ref{Fig8}~(c).
%
%
In addition, our scheme is also robust against decoherence phenomenon causing by the population decay, the depolarization noise arising from the optical pumping effect (see Supplementary Note F for more details) and $50$ Hz noise which originates from the electrical power for a lab (see Supplementary Note G for more details).

\noindent
\textbf{\\DISCUSSION\label{Sec3}}

In summary, we present a general scheme for realizing a quantum double lock-in amplifier via two orthogonal dynamical decoupling sequences: PDD sequence and CP sequence.
This scheme gives a quantum analogue to the classical double lock-in amplifier.
We theoretically derive a general formula for measuring a  AC signal {within strong noise background} via our quantum double lock-in amplifier based upon two quantum mixers under orthogonal modulations.
%
According to our protocol, the complete characteristics of frequency, amplitude and initial phase of a target AC signal can be extracted.
For a weak target signal, one can determine the lock-in point from the symmetry of the combined measurement signal, and then extract the amplitude and even the phase via a fitting procedure.
For a strong target signal, one can extract the target signal from the FFT spectra of the combined measurement signal.

Moreover, we illustrate the realization of a quantum double lock-in amplifier via a five-level double-$\Lambda$ CPT system of $^{87}$Rb as an example.
In the five-level double-$\Lambda$ CPT system, one can measure the common excited-state population to obtain the two measurement signals: $\rho_{55,n}$ and $\tilde\rho_{55,n}$.
%
%
Our scheme shows excellent robustness against finite pulse length and stochastic noise, which is experimentally feasible with state-of-the-art techniques.
Owing to the well-developed techniques in quantum control, various systems can be used to realize the quantum double lock-in amplifier, including Bose condensed atoms~\cite{EDNature2001,BNPRA2011,GPNJP2018,TVNSR2021}, ultracold trapped ions~\cite{MJBPRA2009,PAISR2016,FWMS2021,LDPRAppl2021,KAGScience2021}, nitrogen-vacancy centers~\cite{JHSEL2012,DFPRB2015,CLSR2017,JFBRMP2020,ZQNPJ2022}, and doped spins in semiconductors~\cite{KWCPB2018,PBPRB2018} etc.
Moreover, our protocol may be used for developing practical quantum sensors, such as magnetometers~\cite{MZPRX2021,AKSR2020,MPSR2022,MHPRA2012,GDLScience2010}, atomic clocks~\cite{WFMNature2018,NANJP2019,SDCP2020}, weak-force detectors~\cite{RSNC2017} and noise spectroscopy detector~\cite{JBNP2011,IAPRA2016,SDCP2020}.

\noindent
\textbf{\\METHODS\label{Sec4}}

\noindent
\textbf{\\Calculation of the common excited-state population.}
According to Eqs.~\eqref{Lindblad} and \eqref{Heff}, we can obtain the time-evolution of the density matrix $\rho$ (see Supplementary Note D for more details).
After the adiabatic elimination ($t\gg\frac{1}{\Gamma}$ and $\frac{\partial}{\partial t}\rho_{55}=0$) and assuming the four Rabi frequencies satisfy $\Omega_{j}=\Omega$ and $\{\Omega,\delta\}\ll\Gamma$, we can derive the common excited-state population
\begin{equation}\label{C_main_evolve}
\rho_{55}\approx\frac{\Omega^2}{\Gamma^2}\left(1+\sum_{i,j=1;i\neq j}^{4}\Re{(\rho_{ij})}\right).
\end{equation}
In our consideration, $\rho_{13}$, $\rho_{14}$, $\rho_{23}$ and $\rho_{24}$ are almost 0 before detection, hence the population $\rho_{55}$ can be approximately given as Eq.~\eqref{rho55(0)}.
Due to the two $\Lambda$ configurations $\left\{\ket{1}, \ket{2}, \ket{5}\right\}$ and $\left\{\ket{3}, \ket{4}, \ket{5}\right\}$ have their density matrix elements satisfying $\rho'_{21}=2\rho_{21}$ and $\rho'_{34}=2\rho_{34}$,
we have
\begin{small}
\begin{eqnarray}\label{C_5to3}
\rho_{55}&\approx&\frac{\Omega^2}{2\Gamma^2}\left[\left(1+\Re{(\rho'_{12})}+\Re{(\rho'_{21})}\right)+\left(1+\Re{(\rho'_{34})}+\Re{(\rho'_{43})}\right)\right],\nonumber\\
&=&\frac{\Omega^2}{\Gamma^2}\left(1+2\Re{(\rho_{12})}+2\Re{(\rho_{34})}\right).
\end{eqnarray}
\end{small}
According to the above equation, we find that the five-level double $\Lambda$ configuration can be divided to such two $\Lambda$ configurations.

\noindent
\textbf{Dyson expansion of quantum evolution.}
{In order to simplify the calculation of the time-ordering operator $\hat{\mathcal{T}}$ in Eq.~\eqref{EITOA}, we consider the case of $A t\leq\frac{\pi}{2}$ and utilize the Dyson expansion}
{
	\begin{eqnarray}\label{D_Dyson}
	 &&\hat{U}(t_0,t)=\hat{\mathcal{T}}\exp{\int_{t_0}^{t}\frac{\hat{H}(t')}{i\hbar}dt'}\\\nonumber
    \!&&=\!1\!+\!\!\sum_{n=0}^{\infty}\!\left(\!\frac{-i}{\hbar}\!\right)\!^n\!\!\int_{t_0}^{t}\!\!\!dt_1\!\!\int_{t_0}^{t_1}\!\!\!\!dt_2\!\cdots\!\!\!\int_{t_0}^{t_{n-1}}\!\!\!dt_n
	\hat{H}(\!t_1\!)\!\hat{H}(\!t_2\!)\!\cdots\!\hat{H}(\!t_n\!),
	\end{eqnarray}}
with $\hat{H}(t')=\frac{\hbar}{2}\left[\omega_z(t')\hat{\sigma}_z+\omega_y(t')\hat{\sigma}_y\right]$.
In units of $\hbar=1$, one can obtain
\begin{small}
\begin{eqnarray}\label{D_Dyson_3}
&&\hat{U}(0,t)\approx1-\frac{i}{2}[\phi_z(t)\hat{\sigma}_z+\phi_y(t)\hat{\sigma}_y]\\\nonumber
&&-\frac{1}{4}\!\left[\int_{0}^{t}\!\omega_z(t_1)\phi_z(t_1)dt_1\hat{\sigma}_z^2
+\int_{0}^{t}\omega_y(t_1)\phi_y(t_1)dt_1\hat{\sigma}_y^2\right]\\\nonumber
&&-\frac{1}{4}\left[\int_{0}^{t}\omega_y(t_1)\phi_z(t_1)dt_1\hat{\sigma}_y\hat{\sigma}_z
+\int_{0}^{t}\!\omega_z(t_1)\phi_y(t_1)dt_1\!\hat{\sigma}_z\hat{\sigma}_y\right]\\\nonumber
&&+\mathcal{O}(\phi^3)\\\nonumber
&&\approx 1-\frac{i}{2}[\phi_z(t)\hat{\sigma}_z+\phi_y(t)\hat{\sigma}_y]
	-\frac{1}{8}[\phi_z^2(\!t\!)+\phi_y^2(t)]\\\nonumber
&&-\frac{1}{4}\phi_z(t)\phi_y(t)\hat{\sigma}_y\hat{\sigma}_z-\frac{1}{4}\int_{0}^{t}\omega_z(t_1)\phi_y(t_1)dt_1[\hat{\sigma}_z,\hat{\sigma}_y]\\\nonumber
&&+\mathcal{O}(\phi^3)
\end{eqnarray}
\end{small}
and
\begin{small}
\begin{eqnarray}\label{D_Dyson_3_Total}
&&\hat{U}_{g}(0,t)=\exp{\int_{t_0}^{t}\frac{\hat{H}(t')}{i\hbar}dt'}\\\nonumber
&&\approx 1-\frac{i}{2}\left[\phi_z(t)\hat{\sigma}_z+\phi_y(t)\hat{\sigma}_y\right]\\\nonumber
&&-\frac{1}{8}\left[\phi_z^2(t)\hat{\sigma}_z^2+\phi_y^2(t)\hat{\sigma}_y^2
	 +\phi_y(t)\phi_z(t)\{\hat{\sigma}_y,\hat{\sigma}_z\}\right]
	+\mathcal{O}(\phi^3)\\\nonumber
&&\approx 1-\frac{i}{2}[\phi_z(t)\hat{\sigma}_z+\phi_y(t)\hat{\sigma}_y]
	-\frac{1}{8}[\phi_z^2(t)+\phi_y^2(t)]
	\!+\mathcal{O}(\phi^3).
\end{eqnarray}
\end{small}
Thus, through using  $\{\hat{\sigma}_i,\hat{\sigma}_j\}=\hat{\sigma}_i\hat{\sigma}_j+\hat{\sigma}_j\hat{\sigma}_i=2\delta_{i,j}(i,j=x,y,z)$, the time-ordering operator $\hat{\mathcal{T}}$ in Eq.~\eqref{D_Dyson_3} can be removed when $\phi\leq 2 n \frac{A}{\omega}\leq1$ corresponding to $A t\leq\frac{\pi}{2}$.

\noindent
\textbf{\\DATA AVAILABILITY}

\noindent
Data that support the figures within this paper and other findings of this study are available from the corresponding authors upon reasonable request.

\bibliographystyle{unsrt}

\noindent
\textbf{\\ACKNOWLEDGEMENTS}

\noindent
This work is supported by the National Key Research and Development Program of China (Grant No. 2022YFA1404104), the National Natural Science Foundation of China (Grant No. 12025509), and the Key-Area Research and Development Program of GuangDong Province (Grant No. 2019B030330001).

\noindent
\textbf{\\AUTHOR CONTRIBUTIONS}

\noindent
C.~L. and J.~H. conceived the project. S.~C. and M.~Z. contributed equally to this work and designed the protocol. All authors discussed the results and contributed to compose and revise the manuscript. C.~L. supervised the project.

\noindent
\textbf{\\COMPETING INTERESTS}

\noindent
The authors declare that they have no competing financial interests.

\noindent
\textbf{\\Correspondence}
Correspondence and requests for materials should be addressed to C. Lee~(email: chleecn@szu.edu.cn) and J. Huang~(email: hjiahao@mail2.sysu.edu.cn).

\end{document}